\documentclass[5p,twocolumn]{elsarticle}

\usepackage{graphicx}
\usepackage{amsmath}
\usepackage{amssymb}
\usepackage{bm}
\usepackage{xcolor}
\usepackage{lineno}
\usepackage[version=4]{mhchem}
\usepackage[
    colorlinks=true,
    citecolor=blue,
    linkcolor=blue,
    urlcolor=blue
]{hyperref}
\hypersetup{
    pdftitle={Electrical manipulation of oxygen stoichiometry in multiterminal YBa2Cu3O7-delta junctions},
    pdfauthor={Daniel Stoffels, Caio C. Quaglio-Gomes, Nicolas Lejeune, Emile Fourneau, Pedro Schio, Huidong Li, Lourdes Fabrega, Anna Palau, Maycon Motta, Alejandro V. Silhanek},
    pdfsubject={Materials Today Nano},
    pdfkeywords={YBCO, electromigration, oxygen vacancies, superconductivity, multiterminal devices, defect engineering, nanoelectronics}
}

\journal{Materials Today Nano}

\begin{document}

\begin{frontmatter}

\title{Electrical manipulation of oxygen stoichiometry in multiterminal
\texorpdfstring{YBa$_2$Cu$_3$O$_{7-\delta}$}{YBa2Cu3O7-delta} junctions}

\author[uliege]{Daniel Stoffels\corref{cor1}\fnref{equal}}
\ead{Daniel.Stoffels@uliege.be}

\author[uliege,ufscar]{Caio C. Quaglio-Gomes\corref{cor1}\fnref{equal}}
\ead{caioquaglio@estudante.ufscar.br}

\author[uliege]{Nicolas Lejeune}

\author[uliege]{Emile Fourneau}

\author[lnnano]{Pedro Schio}

\author[icmab]{Huidong Li}

\author[icmab]{Lourdes Fabrega}

\author[icmab]{Anna Palau}

\author[ufscar]{Maycon Motta}

\author[uliege]{Alejandro V. Silhanek\corref{cor1}}
\ead{asilhanek@uliege.be}

\cortext[cor1]{Corresponding authors.}

\fntext[equal]{These authors contributed equally to this work.}

\affiliation[uliege]{
    organization={Experimental Physics of Nanostructured Materials, Q-MAT,
    Department of Physics, Universit\'e de Li\`ege},
    city={Li\`ege},
    postcode={B-4000},
    country={Belgium}
}

\affiliation[ufscar]{
    organization={Departamento de F\'isica, Universidade Federal de S\~ao Carlos},
    city={S\~ao Carlos},
    postcode={13565-905},
    state={SP},
    country={Brazil}
}

\affiliation[lnnano]{
    organization={Brazilian Nanotechnology National Laboratory,
    Brazilian Center for Research in Energy and Materials},
    city={Campinas},
    postcode={13083-100},
    state={SP},
    country={Brazil}
}

\affiliation[icmab]{
    organization={Institut de Ci\`encia de Materials de Barcelona, ICMAB-CSIC},
    addressline={Campus UAB},
    city={Bellaterra},
    postcode={08193},
    country={Spain}
}

\begin{abstract}
Local manipulation of oxygen stoichiometry offers a route to control the electronic properties of complex oxides, yet the selective modification of individual current-carrying branches through oxygen redistribution remains unexplored in multiterminal high-temperature superconducting junctions. In a YBa$_2$Cu$_3$O$_{7-\delta}$ Y-shaped three-terminal device, we demonstrate the possibility to electrically control oxygen vacancy migration on a hand-picked terminal while largely preserving the other two. Oxygen-depleted propagating fronts are directly visualized by the resulting change in optical reflectivity and they are linked to the evolution of the electrical response. The process is highly directional and determined by the polarity of the applied current, allowing for the creation of either a converging or a diverging propagating front from the central node of the Y-shaped device. The associated changes in resistance exhibit relaxation on a timescale of minutes, driven by the vacancy concentration gradient. Effects of oxygen migration are also mapped by Kelvin Probe Force Microscopy and Scanning Laser Microscopy, which probe work-function changes and spatially resolved variations in the superconducting transition, respectively. Notably, the $T_c$ contrast revealed by the latter provides a quantitative handle on the underlying oxygen content, enabling direct visualization of oxygen redistribution. Finite-element modeling and nanoprobe X-ray diffraction qualitatively reproduce the observed vacancy redistribution. These results establish a post-fabrication route to locally tune properties of superconducting multiterminal devices such as nanocryotrons, yTron, and tunable weak links.
\end{abstract}


\end{frontmatter}

\section{Introduction}
Over the past several decades, the phenomenon of current-induced atom migration has undergone a paradigm shift: originally viewed as a major reliability issue that limited the operational lifetime of integrated circuits \cite{Lloyd_1997}, it has evolved into a controllable and cost-effective strategy for nanoscale fabrication \cite{Hoffmann-Vogel}. The technique has been exploited across a diverse range of applications, including formation of nanogaps for single-molecule electronic junctions \cite{Mahapatro,Strachan}, realization of plasmonic nanoantennas \cite{Gurunarayanan2017}, removal of surface contaminants in two-dimensional materials \cite{Moser}, creation of quantum point contacts \cite{Hoffman}, modulation of superconducting weak links \cite{Baumans2016,Lombardo,Keijers,Collienne-2021}, fabrication of metallic nanowires \cite{KIMURA}, memristors \cite{Lombardo_2019} and nanoheaters \cite{BlancoAlvarez2019}. The driving mechanism originates from the interplay between momentum transfer from charge carriers to lattice ions and the action of the electric field on the ions, a phenomenon known as electromigration (EM). Effective implementation requires the simultaneous presence of high current densities, elevated local temperatures, and sufficiently weak atomic binding energies, conditions readily satisfied in a broad class of metallic systems.

An especially intriguing case arises in complex oxides such as the high-temperature superconductor YBa$_2$Cu$_3$O$_{7-\delta}$ (YBCO), in which charge carriers are predominantly hole-like. In addition, the ceramic nature of the system leads to relatively high resistivity values, implying that the direct force exerted by the electric field on mobile species cannot be disregarded \cite{Jacobs}. Charge transport occurs preferentially along well-defined pathways associated with the Cu--O chains and CuO$_2$ planes \cite{Friedmann}. Notably, the O(1) sites located within the chains exhibit enhanced mobility and become susceptible to displacement \cite{Rothman}. Under such conditions, the oxygen stoichiometry can be modified by applying an electric current without inducing significant structural degradation. This effect has been exploited both to reconstruct the phase diagram of critical temperature as a function of oxygen content \cite{Trabaldo} and to achieve in situ tuning of nanowire-based YBCO SQUIDs \cite{Trabaldo_2021}. Interestingly, these modifications can be undone, provided they involve a reversible oxygen motion within a moderate atomic displacement regime \cite{Stoffels}. Most of these works have focused on transport bridges with a well defined constriction, ensuring the spatial localization of the EM due to current crowding effects.

The extension of the above approach from single constrictions to multiterminal devices has been investigated in Nb-based circuits. In Ref.~\cite{collienne2022}, it was demonstrated that in three-terminal junctions, the properties of one terminal can be selectively modified without affecting the remaining two. Similar behavior was observed in systems comprising multiple terminals, where the spatial extent of the modified region was shown to depend sensitively on the terminal geometry \cite{marinkovic_elijah_2022}. From this perspective, potential applications of the EM process in three-terminal devices can be found in the domain of superconducting electronics, such as the nanocryotron proposed in Refs.~\cite{McCaughan2014,Foster}, the yTron proposed as a sensor and readout of current-flow in a superconductor \cite{mccaughan_using_2016}, or in tunable superconducting weak links by injecting a normal current into the junction \cite{volkov_new_1995,baselmans_direct_2002,winik_local_2018}.

In this work, we investigate three-terminal devices made of YBCO and demonstrate the selectivity of the EM process, similar to what has already been reported in Nb but without involving major structural changes. The displacement of oxygen vacancies is captured by multimodal characterization, with local changes detected by optical reflectivity, Kelvin Probe Force Microscopy (KPFM), and Scanning Laser Microscopy (SLM). Depending on whether the current density points away from or towards the central node of the Y-shaped junction, oxygen vacancies can be correspondingly moved away from or concentrated at the center, highlighting the important role of current polarity. We show that the retention of the EM-induced high-resistance state is limited by a relaxation process driven by oxygen concentration gradients. The relaxation time follows an Arrhenius temperature dependence, consistent with a thermally activated diffusion process. This allows us to extract an activation energy and compare it with oxygen diffusion barriers reported in previous studies. SLM images enable the mapping of oxygen content by locally assessing the superconducting critical temperature. Experimental findings are qualitatively reproduced by finite element modeling and permit identification of the prevailing mechanisms at play during EM and relaxation processes.

\begin{figure*}[ht]
    \centering
    \includegraphics[width=\linewidth]{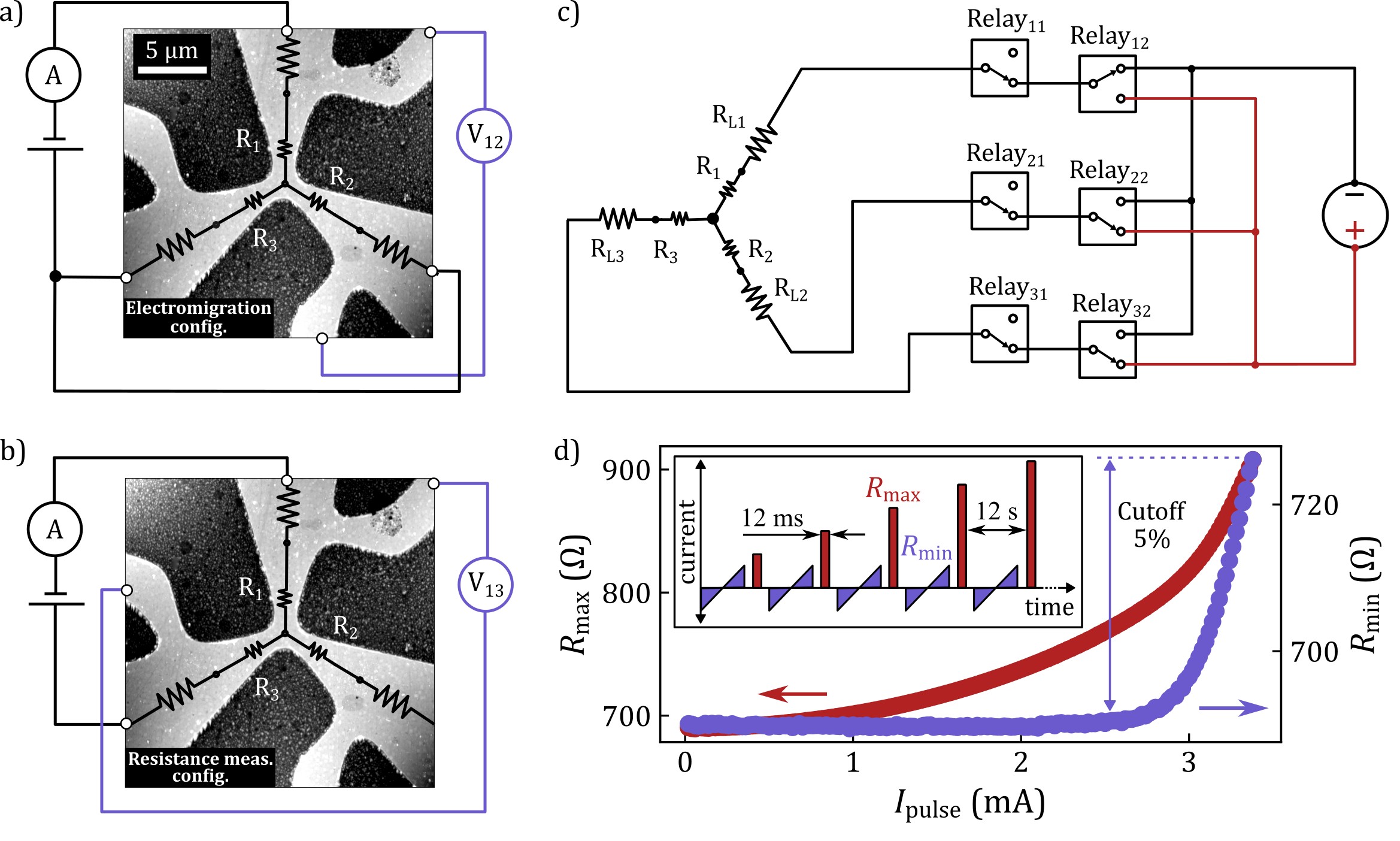}
    \caption{
        \textbf{Methods:} (a) AFM image overlaid with the schematic electrical connections for EM applied to terminal 1 with negative polarity. (b) Same AFM image with the electrical connections for resistance measurement of $R_{13}=R_{1}+R_{3}$ between terminals 1 and 3. (c) Equivalent circuit and relay wiring used to automate configuration switching during electrical measurements. Each terminal is modeled by a resistor $R_i$, while the additional resistor $R_{Li}$ represents the series resistance of the remaining material outside the region between the voltage probes, i.e., the ohmic contribution from areas farther from the central node. (d) Experimental electrical response during a representative EM cycle. The left y-axis shows the resistance $R_{\mathrm{max}}$, measured during each current pulse, while the right y-axis shows the resistance, $R_{\mathrm{min}}$, obtained during low-current probing windows between successive pulses. Inset: Schematic of the EM pulse protocol, consisting of linearly increasing amplitude current pulses of 12 ms 
        duration to drive atomic diffusion in the selected terminal, followed by a 12 s probing window comprising low-current ramps of both polarities to measure the sample resistance with negligible Joule heating.
    }
    \label{fig:setup}
\end{figure*}
 
\section{Methods}

\subsection{Sample preparation}
The devices studied in this work were fabricated from 50 nm-thick epitaxial YBCO thin films grown on \ce{SrTiO3} (STO) substrates by Pulsed Laser Deposition (PLD), with similar deposition parameters as those reported in Ref.~\cite{Lejeune2025}. The films were patterned into a three-terminal geometry using photolithography followed by ion-beam etching. The narrowest section of the device measures approximately 2~\textmu m, as depicted in the Atomic Force Microscopy (AFM) image in Fig.~\ref{fig:setup}(a,b). Ohmic contacts were obtained by sputter deposition of 50 nm gold pads after ion milling. Several devices with superconducting critical temperatures spanning from $83$ to $89$~K were measured, all yielding similar EM behavior.

\subsection{Electric measurements}

Each device can be modeled as three resistors, $R_1$, $R_2$, and $R_3$, connected to a central node, allowing extraction of the individual contributions of each terminal from the pairwise resistances $R_{ij}=R_i+R_j$, for $i,j=1,2,3$ and $i \neq j$. A schematic representation of the setup can be found in Fig.~\ref{fig:setup}. The electrical configurations used for EM and for assessment of pairwise resistances are shown in Fig.~\ref{fig:setup}(a) and Fig.~\ref{fig:setup}(b), respectively. Figure~\ref{fig:setup}(c) illustrates the equivalent circuit together with the relay-controlled wiring used to automate switching between measurement modes. Additional series resistors $R_{Li}$ account for the two-wire resistance of the full measurement circuitry.

Electromigration is performed by concentrating the highest current density in a selected terminal. Two current polarities are defined: in the negative polarity, the target terminal is connected to the current drain while the other two terminals are connected to the current source (as in Fig.~\ref{fig:setup}(a)), whereas in the positive polarity, the target terminal is connected to the source and the remaining terminals to the drain. To achieve controlled oxygen migration, a pulsed protocol is employed, as schematically illustrated in the inset of Fig.~\ref{fig:setup}(d). A sequence of 12~ms current pulses with linearly increasing amplitude in 0.02~mA increments is applied to drive EM in the selected terminal. Between successive pulses, a 12~s probing window is used to acquire low-current $I$--$V$ curves between $-100$~\textmu A and $100$~\textmu A, allowing resistance measurements with negligible Joule heating ($\mathcal{P}_{\mathrm{probe}} \approx 10$~\textmu W $\ll \mathcal{P}_{\mathrm{EM}}$). Figure~\ref{fig:setup}(d) shows the electrical response during a representative EM cycle obtained using this protocol. The resistance measured during each current pulse is denoted by $R_{\mathrm{max}}$, which is affected by Joule heating, while the resistance extracted from the low-current probing window is denoted by $R_{\mathrm{min}}$. This protocol enables real-time monitoring of the sample response during EM, providing a direct means to follow the evolution of the terminal resistances and to determine when to interrupt the electrical stress. Each EM run consists of a sequence of pulses terminated when a cutoff resistance is reached, typically defined as a 5\% increase relative to the initial value of $R_{\mathrm{min}}$.

\subsection{Far-field optical microscopy}
The dielectric tensor of YBCO depends on the local oxygen content, leading to a reflectance that varies with oxygen stoichiometry and light polarization, as thoroughly characterized in the 2--5~eV energy range in Ref.~\cite{kircher1991}. Consequently, local variations in oxygen stoichiometry produce measurable optical contrast, with deoxygenated regions exhibiting higher reflectivity and thereby enabling direct visualization of oxygen migration \cite{Moeckly,marinkovic_direct_2020,Liu,Biancardi2026}. To monitor the oxygen depleted regions in situ, we employed conventional far-field optical microscopy, acquiring images of the YBCO devices during or after EM experiments. The optical setup consists of an Olympus 50$\times$ objective combined with built-in 2$\times$ magnification, a non-polarized LED white light source combined with a green bandpass filter ($\lambda = 550 \pm 50$~nm), a linear polarizer, and a Retiga 4000R 12-bit CCD camera. Prior to selecting this configuration, different illumination wavelengths and both polarized and unpolarized conditions were tested, and the EM effects were consistently observed across all tested modes, however the chosen configuration provided the highest optical contrast. Image processing was performed using the software Fiji \cite{schindelin_fiji_2012}: outliers were masked, and background contributions were removed by subtracting a reference image acquired prior to the onset of EM.

\subsection{Kelvin-probe force microscopy}
KPFM enables nanoscale mapping of surface potential by measuring the local work function difference between a conductive tip and the sample surface \cite{KPFM,MELITZ20111}. In this technique, an alternating voltage applied to the cantilever generates an oscillating electrostatic force between the tip and the surface. The resulting signal, detected by a lock-in amplifier, is compensated by applying a DC bias to the tip that nullifies the electrostatic interaction. The value of this compensating voltage corresponds to the contact potential difference ($V_\mathrm{CPD}$) and thus reflects local variations in the work function. KPFM has been effectively employed to visualize oxygen-content variations in YBCO \cite{Trabaldo}. In the present study, measurements were carried out using a Nanosurf DriveAFM system equipped with a Nanosensors PPP-EFM conductive probe with a PtIr$_5$ metal coating, allowing high-resolution detection of surface potential contrasts associated with EM-induced oxygen redistribution.

\subsection{Scanning Laser Microscopy}
\label{sec:methods_slm}
SLM was employed to spatially resolve the local superconducting transition of an electromigrated device. To achieve this, the sample surface is scanned with a 532 nm laser, focused to a ${1}/{e^2}$-spot size of 1.5~\textmu m, and modulated by an optical chopper at a frequency of 188 Hz. During the measurements, the device was biased with a low current of 10 \textmu A while the voltage response was monitored with a lock-in amplifier. The periodic laser heating, confined to a region a few times larger than the spot size \cite{slm, Carslaw_1959, Lejeune2025}, induces a local resistivity modulation, generating a spatial map $V(x,y)$. This procedure was repeated over a temperature sweep from 30 K (sample is superconducting everywhere) to 95 K (sample is entirely in the normal state) in 0.5 K steps to obtain a map of $dR/dT\ (x,y)$, which is then numerically integrated with respect to $T$ to reconstruct the local $R(T)$ behavior and extract a point-by-point estimate of $T_c$ across the device.

\subsection{Finite-element modeling}
\label{sec:methods_sim}
The spatial and temporal evolution of oxygen and temperature in the devices was simulated using the software suite COMSOL Multiphysics assuming a two-dimensional YBCO film with a virtual thickness of $50~\mathrm{nm}$. The film was placed on a disk of substrate with a diameter of $60$~\textmu m and a thickness of $5$~\textmu m. The model couples the electrical current distribution, Joule heating, and oxygen redistribution in the YBCO layer. Heat evacuation was only allowed through the bottom surface of the substrate, which was kept at 300~K, while heat losses through the top surface were neglected. Electrical boundary conditions were imposed to reproduce the connection scenario depicted in Fig.~\ref{fig:setup}(a) for the selective EM of one terminal. This finite-element framework is a further development of the one reported in Refs.~\cite{collienne2022b, Stoffels}, with modifications introduced to reproduce the present pulsing sequence, include relaxation between successive pulses, and prevent unphysical oxygen accumulation beyond full oxygenation.
The oxygen state was described by a normalized oxygen content $X_\mathrm{O}$ following the convention used in Ref.~\cite{collienne2022b}. This dimensionless variable maps the mobile oxygen content onto the interval $0 \leq X_\mathrm{O} \leq 1$, where $X_\mathrm{O}=0$ corresponds to the tetragonal, strongly oxygen-depleted state and $X_\mathrm{O}=1$ to full oxygenation of the mobile oxygen sublattice. In this description, only the oxygen content in excess of the tetragonal state is allowed to participate in diffusion and EM.
The resistivity of the YBCO film was calculated from an oxygen-dependent resistivity law based on the data provided in Ref.~\cite{Semba2001}, with the initial oxygen concentration adjusted through the experimentally measured $T_\mathrm{c}$. For numerical stability, the tabulated resistivity dependence from Ref.~\cite{Semba2001} was implemented using a spline interpolation. Finally, the absolute four-wire resistance was calibrated by multiplying the resistivity law by a constant factor chosen to reproduce the pristine device resistance.

The simulated pulsing sequence followed the same structure as the experimental protocol: First, the resistance state was evaluated at low readout current, without Joule heating or oxygen motion. Then, a current pulse of duration $12~\mathrm{ms}$ was applied, during which the electrical, thermal, and oxygen-transport equations were solved in a coupled manner. After each pulse, the system was allowed to relax for $12~\mathrm{s}$ at room temperature, considering concentration-driven oxygen diffusion only. The resulting oxygen profile was used as the initial condition for the next pulse amplitude, in which the pulse current was increased by 0.02 mA. Thus, the simulated sequence used exactly the same pulse-current increment, pulse duration, and inter-pulse relaxation time as the experiment.

Oxygen redistribution was modeled by diffusion and EM according to
\begin{equation}
    \frac{\partial X_\mathrm{O}}{\partial t}
    =
    \nabla \cdot
    \left[
        D(T)\nabla X_\mathrm{O}
        -
        D(T)\frac{Z^\ast F}{RT}
        X_\mathrm{O}\left(1-X_\mathrm{O}\right)\mathbf{E}
    \right],
    \label{eq:oxygen_transport}
\end{equation}
where $\mathbf{E}$ is the local electric field, $Z^\ast$ is the effective EM charge number, $F$ is the Faraday constant, $R$ is the gas constant, and $T$ is the local temperature. The diffusion coefficient was written in Arrhenius form,
\begin{equation}
    D(T)=D_0\exp\left(-\frac{E_a}{k_\mathrm{B}T}\right),
    \label{eq:oxygen_diffusion}
\end{equation}
using $D_0 = 1.4 \cdot 10^{-8}$ m$^2$/s from Ref.~\cite{Rothman} and an activation energy $E_a=0.5~\mathrm{eV}$, supported by the experimental findings in Sec.~\ref{sec:relax}. The factor $\left(1-X_\mathrm{O}\right)$ in Eq.~(\ref{eq:oxygen_transport}) suppresses oxygen accumulation as $X_\mathrm{O}$ approaches 1, preventing unphysical overdoping in the normalized concentration description.

Mesh convergence was checked by repeating the full pulsing sequence for several mesh densities and comparing the resulting $R_\mathrm{min}(I_\mathrm{pulse})$ curves and oxygen profiles. The mesh was chosen such that further refinement did not lead to appreciable changes in the pulse current at which oxygen redistribution sets in, nor in the final oxygen profiles.

\section{Results and Discussion}
\begin{figure*}[ht]
    \centering
    \includegraphics[width=0.95\linewidth]{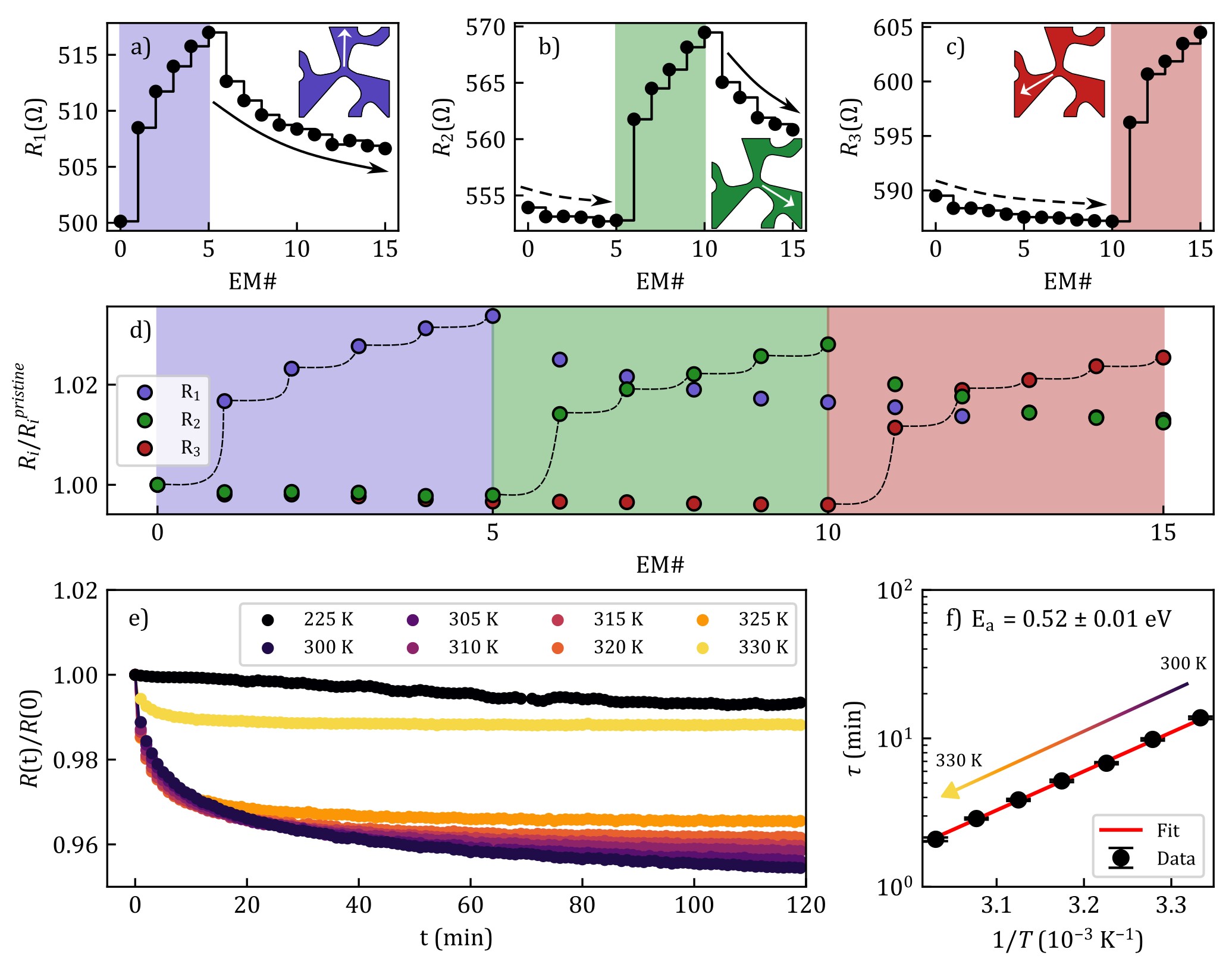}
    \caption{\textbf{Results of 15 EM runs highlighting the method’s selectivity:} During runs 1–5, terminal R$_1$ was targeted; runs 6–10 targeted R$_2$; and runs 11–15 targeted R$_3$. Panels (a)–(c) show the evolution of the resistances of terminals R$_1$, R$_2$, and R$_3$, respectively, measured after each EM run. Panel (d) displays the resistances normalized to their initial values. Panel (e) presents the resistance relaxation as a function of time following eight 5\% EM runs performed at different temperatures and monitored for 2 h. Stretched-exponential fits yielded temperature-dependent relaxation time constants with an average stretching exponent of $\beta = 0.455 \pm 0.004$; almost no relaxation was observed at 225 K within the same time window. Panel (f) shows the linearized Arrhenius plot of the relaxation time constant as a function of inverse temperature, from which an activation energy of $E_a = 0.52 \pm 0.01$ eV was obtained.}
    \label{fig:EMs}
\end{figure*}

\subsection{Targeted electromigration}
Multiterminal devices, and in particular the Y-shaped geometry, allow us to illustrate the ability to select the terminal where atomic displacement will occur by applying the appropriate electrical configuration. Additionally, this design enables the determination of individual terminal resistances $R_i$, which are otherwise experimentally inaccessible through direct measurement. Experimental evidence of the targeted EM process is shown in Fig.~\ref{fig:EMs}, presenting the final resistance measured when the stop criterion is met after each of five EM runs applied successively to each terminal under negative polarity. The initial resistances of the three branches range from 500~$\Omega$ to 590~$\Omega$, reflecting slight differences in geometry. Each EM run was stopped after a 5\% increase in $R_{\mathrm{min}}$, as previously illustrated in Fig.~\ref{fig:setup}(d). 

Under negative current polarity, oxygen vacancies are expected to drift away from the central node and accumulate downstream along the driven terminal, leading to an increase in its resistance. This behavior can be interpreted in terms of the effective valence $Z^{*}$, which describes the net electromigration force acting on the migrating species \cite{Hoffmann-Vogel}. Since oxygen vacancies drift along the direction of the conventional current (i.e., the direction of positive charge flow), the corresponding O$^{2-}$ ions move in the opposite direction. Within our sign convention, this corresponds to an effective electromigration charge $Z^{*}<0$ for oxygen under the present experimental conditions. Importantly, $Z^{*}$ should not be identified with the nominal ionic charge. As discussed by Truchly et al.~\cite{Truchly2016}, the effective electromigration force contains contributions from both the partially screened ionic charge and momentum transfer from the current carriers (the ``charge-wind'' force). In hole-doped YBCO, the balance between these contributions depends on the local conductivity and carrier concentration and may therefore modify both the magnitude and, in principle, the sign of $Z^{*}$.

Figures~\ref{fig:EMs}(a--c) show the sequential EM of terminals $R_1$, $R_2$, and $R_3$. For the sake of comparison, Fig.~\ref{fig:EMs}(d) shows the same data with normalized resistance values. The most salient feature of this figure is the high degree of selectivity: when one terminal undergoes EM, only its resistance changes significantly. Nevertheless, during electrical stress to terminal $i$, a slight decrease in resistance is observed in the neighboring terminals, $j \neq i$ as indicated by the black dashed arrows in Fig.~\ref{fig:EMs}(b--c). For previously untargeted neighboring terminals, this decrease is attributed to oxygen replenishment as oxygen atoms migrate away from terminal $i$ toward the central node. In contrast, when a neighboring terminal has already undergone EM, the observed resistance decrease reflects the combined effects of oxygen replenishment and a relaxation process, shown by the black solid arrows in Fig.~\ref{fig:EMs}(a--b), which is discussed in detail in the following section.

\subsection{Relaxation process}
\label{sec:relax}
Note that although each EM step was intended to produce a similar resistance increase, the observed increments progressively decrease due to cumulative relaxation processes. Indeed, each EM step accentuates the gradients of vacancy concentration, which are the driving force for the relaxation process. This relaxation occurs immediately after EM, when the terminal is probed without any electrical stress. To further investigate this effect, we monitored the post-EM resistance for EM runs with a 5\% cutoff performed between 225 K and 330 K, as depicted in Fig.~\ref{fig:EMs}(e). A single new device was used for the relaxation measurements, as relaxation was found to be history-independent (see Supplementary Material S1), and the temperature sequence was randomized to avoid systematic ordering effects. At the lowest temperature, $225$~K, only a very weak relaxation was observed within a 120--min monitoring window. At higher temperatures, the relaxation process becomes more apparent and follows a stretched exponential dependence,

\begin{equation}
    R(t)=R_{\infty}+\Delta R\exp{\left[-\left(\frac{t-t_{0}}{\tau} \right)^{\beta }\right]},
\end{equation} 
 
\noindent with time constants $\tau$ ranging from 13.75 to 2.08 minutes between 300 K and 330 K. The obtained averaged stretched exponent $\beta~=~0.455\pm 0.004$ is consistent with similar relaxation processes observed in YBCO under photoexcitation \cite{Lejeune2025}. This functional form is characteristic of disordered systems and phases, in which distinct local environments with distinct energy barriers yield a distribution of relaxation times. As shown in Fig.~\ref{fig:EMs}(f), plotting $\tau$ versus $1/T$ and applying an Arrhenius fit, we obtain an activation energy of $E_a = 0.52 \pm 0.01$ eV, within the range of oxygen diffusion energies reported for YBCO \cite{Rothman1989,Veal1990}. It is worth mentioning that similar exponential relaxation behavior after electrical stress has been reported in Al structures, albeit with much longer characteristic times \cite{Loyd-relaxation}. Additionally, negligible relaxation was observed in three-terminal Nb samples \cite{collienne2022}, testifying to the high mobility of oxygen vacancies in YBCO. 

\subsection{Visualization of the oxygen-depleted regions}
\subsubsection{Optical reflectivity}
\begin{figure*}[ht]
    \centering
    \includegraphics[width=\linewidth]{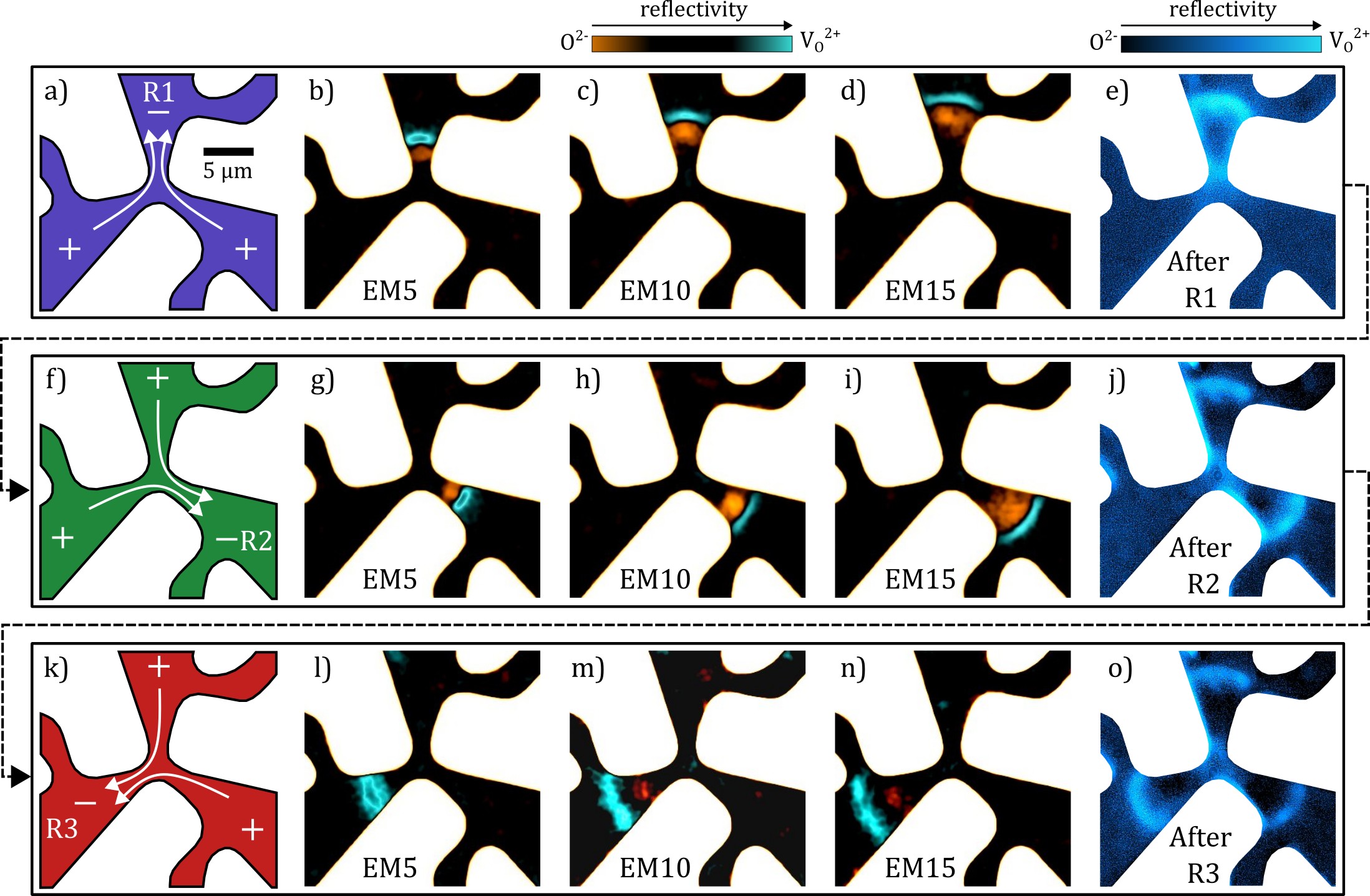}
    \caption{
        \textbf{Results obtained from optical microscopy collected during 15 successive EM runs performed on each terminal of the same device:} Panels (a-e), (f-j), and (k-o) correspond to EM of terminals R1, R2, and R3, respectively. In each group, the first panel illustrates the electrical configuration used to selectively drive EM, while the following three panels show differential optical micrographs acquired after 5, 10, and 15 runs. In these images, the substrate appears in white, the YBCO film in black, while oxygen-depleted regions (with higher concentration of oxygen vacancies, $\mathrm{V_{O}}^{2+}$), characterized by increased optical reflectivity, are highlighted in light blue. The orange features arise from the differential imaging procedure and represent the motion of the deoxygenation front, leaving behind a trail of increased oxygen content (i.e., a higher concentration of oxygen ions, $\mathrm{O}^{2-}$). The last panel in each group shows the accumulated effect of all EM runs, where overall reflectivity changes are emphasized with a blue scale.
    }
    \label{fig:optical}
\end{figure*}

A more compelling piece of evidence for the targeted EM process comes from direct optical visualization of the oxygen-deficient zone, as manifested by its change in reflectivity \cite{Moeckly}. Fig.~\ref{fig:optical} summarizes a selected set of snapshots of the optical response for sequences of 15 EM runs applied to each terminal. The complete image sequence is provided as a video in Supplementary Video 1. Again, each run was halted upon reaching a 5\% increase in $R_{\mathrm{min}}$, and each subsequent run was initiated from the final current value of the preceding one. In the first column, the target terminal and the applied current polarity are indicated, with the conventional current direction represented by white arrows. The three central columns show incremental differential optical images after 5, 10, and 15 runs, each obtained by subtracting the image acquired immediately before the corresponding run. In these images, light blue regions indicate zones where oxygen depletion takes place, whereas orange trails result from image differentiation and oxygen accumulation behind the deoxygenation front. These images provide unambiguous evidence that oxygen vacancy motion is highly directional, consistent with an EM-governed mechanism and minimal thermomigration effects. The separation between the two colored lobes is linked to the increase in resistance during the corresponding EM step, with a greater increase in resistance leading to more widely spaced lobes. The rightmost column of Fig.~\ref{fig:optical} shows the cumulative optical changes induced by 15 EM runs on each terminal, obtained by subtracting the pristine image from the corresponding final image. It is worth noting that the rims along the sample's edges show a higher degree of oxygen loss.

\subsubsection{Work function}
\begin{figure}[ht]
    \centering
    \includegraphics[width=\linewidth]{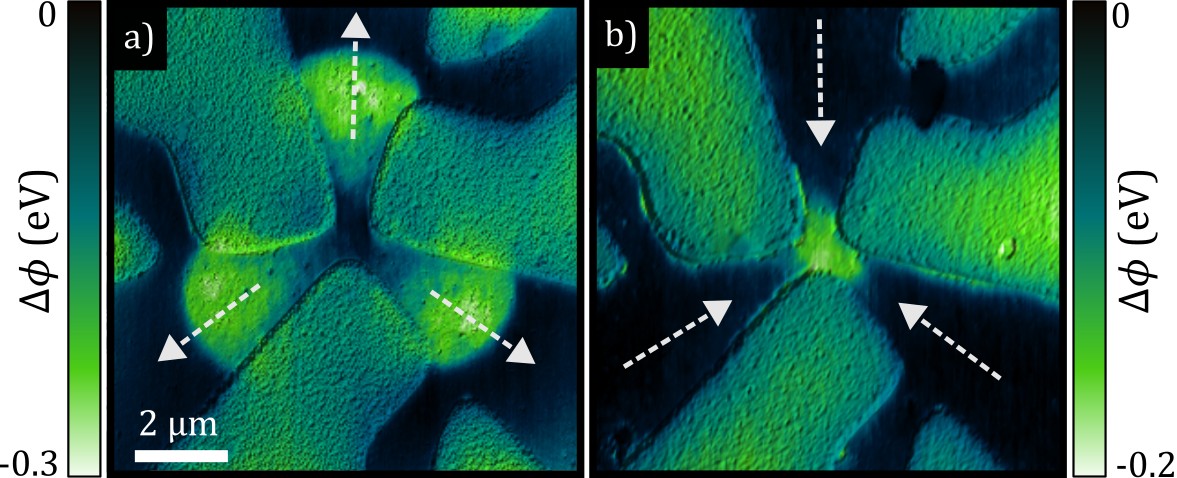}
    \caption{
        \textbf{Kelvin Probe Force Microscopy:} (a) KPFM map of the YBCO device after EM under negative polarity (same device as Fig.~\ref{fig:optical}), showing work function variations ($\Delta \phi$) associated with oxygen redistribution. (b) KPFM map of a second device after five EM cycles under positive polarity applied to each terminal. White arrows indicate the inferred direction of vacancy migration. The zero references correspond to pristine regions far from the center of the devices.
    }
    \label{fig:kpfm_sem}
\end{figure}

For a semi-quantitative and spatially resolved analysis of EM effects, we employed KPFM, which is sensitive to oxygen redistribution through the doping dependence of the YBCO work function. Figure~\ref{fig:kpfm_sem}(a) shows the KPFM map of the device after 15 EM runs under negative polarity (same as in Fig.~\ref{fig:optical}). The surface potential exhibits a gradual variation away from the central node, reflecting the redistribution of oxygen vacancies induced by EM. The zero reference is defined on a pristine YBCO region far from the center of the device. The measured work function difference, defined as $\Delta \phi = \phi_\mathrm{YBCO} - \phi_\mathrm{tip} = -eV_\mathrm{CPD}$, is approximately $0.30$~eV between the potential minimum and the pristine region, consistent with previous comparisons between underdoped and slightly overdoped YBCO \cite{Trabaldo}. A comparison between the KPFM and optical images of the same device (Fig.~\ref{fig:optical}) reveals that the spatial extent of the affected regions differs, with the work function signal being more spread out and less confined than the optical signal. We attribute this difference to a slow post-EM relaxation of the oxygen distribution. Indeed, since the KPFM measurements were performed approximately two weeks after the optical characterization, they likely reflect continued oxygen redistribution during this time. Still, the dark contrast of the central node hints at oxygen movement inwards when driving a vacancy front outwards. 

Figure~\ref{fig:kpfm_sem}(b) shows the KPFM map of a second device after five successive EM cycles under positive polarity performed for each terminal. In contrast to the negative-polarity case (panel (a)), the surface potential distribution indicates vacancy accumulation at the central node. In this case, the maximum work function difference is approximately -0.2~eV, smaller than that observed for negative polarity, which is attributed to the smaller accumulated EM dose resulting from the lower number of pulsing cycles. Additional experimental results and finite-element simulations for the positive-polarity case are provided in the Supplementary Material S2.
Together, the two KPFM maps provide direct and unambiguous evidence that reversing the current polarity reverses the direction of oxygen vacancy migration, demonstrating the ability to create either diverging or converging oxygen-vacancy fronts through EM.

\begin{figure}[ht]
    \centering
    \includegraphics[width=\linewidth]{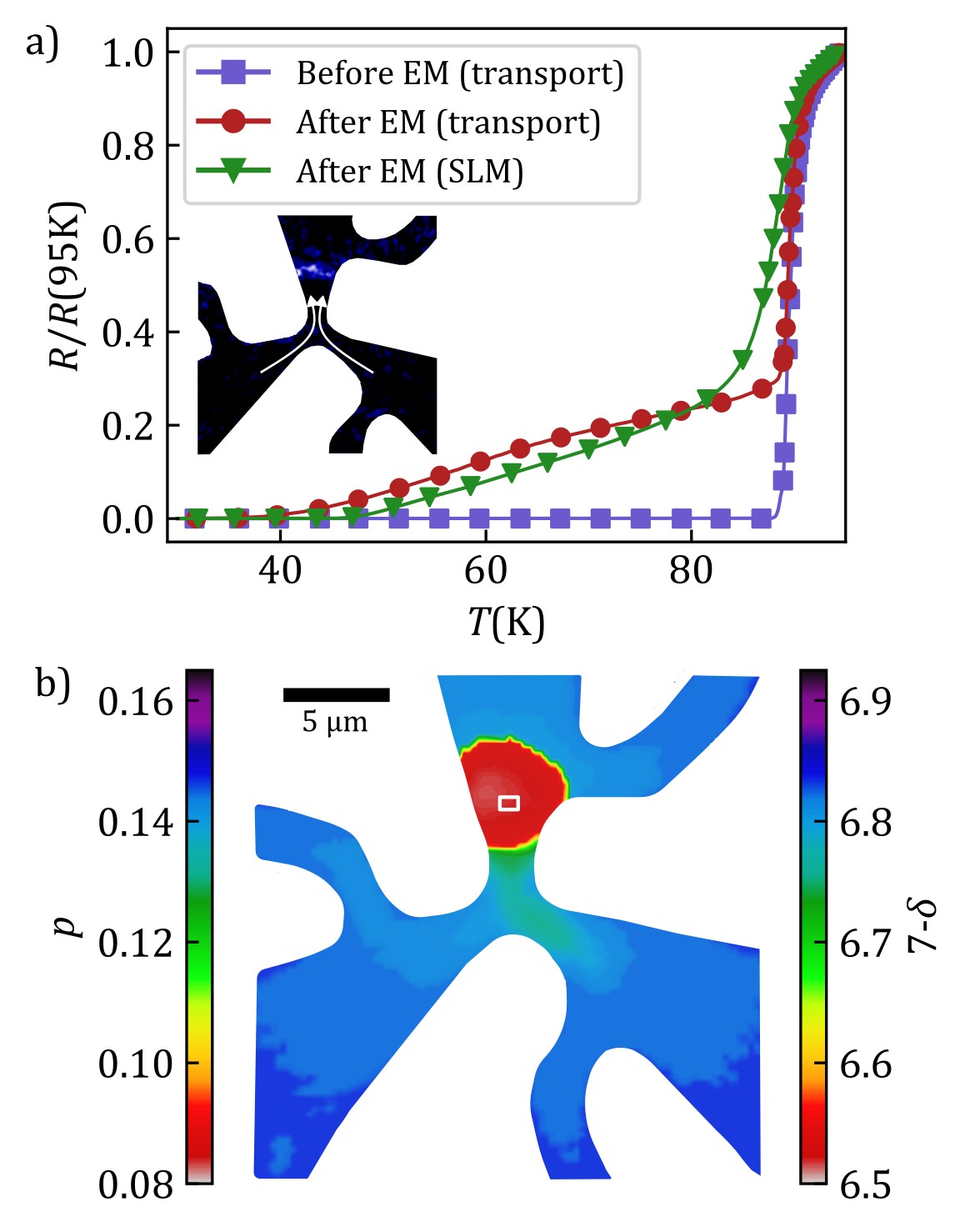}
    \caption{
        \textbf{Scanning Laser Microscopy:} 
        (a) Normalized $R(T)$-curves of a three-terminal device measured before (circles) and after (squares) EM. A reconstructed $R(T)$-curve, obtained via SLM during a temperature sweep, is also shown (triangles). Inset: false-colored optical image of the device after EM, highlighting the oxygen vacancy front and the current polarity. (b) Hole doping ($p$) map derived from the $T_c$ map obtained via SLM, using the empirical relation of Ref.~\cite{Presland1991} (left colorbar) and oxygen content ($7-\delta$) map based on the experimental $p$ vs. oxygen content relation reported in Ref.~\cite{Liang2006} (right colorbar). A white rectangle on the map in panel (b) indicates the region where the SLM $R(T)$-curve of panel (a) was averaged.
    }
    \label{fig:slm}
\end{figure}

\subsection{Mapping of the oxygen content}
Quantitative mapping of the oxygen content can be achieved through scanning laser microscopy. The technique detailed in Sec.~\ref{sec:methods_slm} allows the reconstruction of spatially resolved $R(T)$ curves and  has, to our knowledge, not been previously employed for characterizing electromigrated samples. As an example, Fig.~\ref{fig:slm}(a) shows the $R(T)$-curves reconstructed using this technique, averaged over a region of 400 nm $\times$ 600 nm at the center of the oxygen-depleted region. For the sake of comparison, the same plot includes the $R(T)$-curves before and after EM as measured via electrical transport measurements. The foot in the $R(T)$ at the superconducting transition reveals a region with lower $T_c$ along the affected current terminal, as reported previously \cite{collienne2022b}. Applying this technique pixel-by-pixel, we can obtain a $T_c$ map (upon significant deviation from zero resistance), from which we derive maps of hole doping and oxygen content, as shown in Fig.~\ref{fig:slm}(b). The left colorbar corresponds to the hole concentration ($p$) per copper atom in the CuO$_2$ planes calculated using the empirical Presland relation \cite{Presland1991}, while the right colorbar displays the oxygen content $(7-\delta)$ derived from the experimental correlation between $p$ and oxygen stoichiometry reported by Liang et al.~\cite{Liang2006}. For the evaluation of $p$, the maximum critical temperature $T_{c,\mathrm{max}}~=~89.3$ K was determined as the point of maximum slope of the pristine, conventionally measured $R(T)$ curve.

Besides revealing localized effects, the oxygen map illustrates how deoxygenation propagates along the electromigrated terminal. Compared to the differential optical image shown as an inset in Fig.~\ref{fig:slm}(a), the oxygen-depleted region appears larger and spreads radially, while traces of less-deoxygenated YBCO remain toward the center of the device. The apparent enlargement can be attributed to a larger effective laser spot size due to the heat diffusion. This spreading of the probing spot induces voltage signals in regions that are not actively electromigrated, and therefore, the obtained signal results from a convolution product. A more thorough description can be found in Ref.~\cite{Lejeune2025, lejeune2026_time}. This convolution effect also explains the fact that the estimated hole doping varies approximately from $p \sim 0.08$ to $p \sim 0.15$ when moving from the most strongly deoxygenated areas to regions closer to pristine material, corresponding to an oxygen content ranging from roughly 6.5 to 6.85, a value slightly below the commonly accepted optimal doping.

\subsection{Modeling the electromigration process}
\begin{figure*}[ht]
    \centering
    \includegraphics[width=1\linewidth]{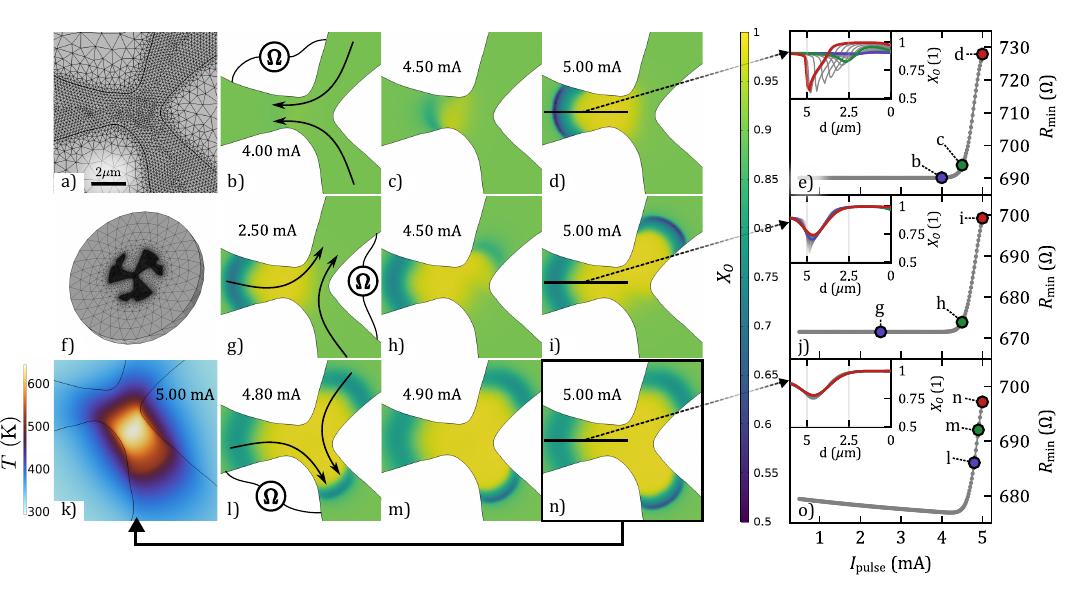}
    \caption{
        \textbf{Overview of the simulated multiterminal pulsing sequence.}
        (a) Close-up of the finite-element geometry and mesh in the constriction region.
        (b--d) Oxygen-content maps $X_\mathrm{O}$ for the first wiring configuration, with pulse amplitudes $I_\mathrm{pulse}=4.00$, $4.50$, and $5.00~\mathrm{mA}$.
        The sketch in (b) indicates the current-flow direction and the terminal pair used for the resistance measurement.
        The black line in (d) marks the position of the line profile shown in the inset of (e).
        (e) Corresponding evolution of the measured resistance $R_\mathrm{min}$ as a function of $I_\mathrm{pulse}$.
        Colored markers indicate the pulse amplitudes shown in (b--d), and the inset shows the associated line profiles of $X_\mathrm{O}$.
        (f) Full simulation domain and mesh, including the disk-shaped substrate region.
        (g--i) Oxygen-content maps for the second wiring configuration, starting from the final state of (d), at $I_\mathrm{pulse}=2.50$, $4.50$, and $5.00~\mathrm{mA}$.
        The sketch in (g) indicates the current-flow direction and resistance-measurement terminals for this configuration.
        (j) Corresponding $R_\mathrm{min}(I_\mathrm{pulse})$ curve and oxygen line profiles, with colored markers matching panels (g--i).
        (k) Temperature map $T$ for the configuration shown in (n).
        (l--n) Oxygen-content maps for the third wiring configuration, at $I_\mathrm{pulse}=4.80$, $4.90$, and $5.00~\mathrm{mA}$.
        The sketch in (l) indicates the current-flow direction and measurement terminals.
        (o) Corresponding $R_\mathrm{min}(I_\mathrm{pulse})$ curve and oxygen line profiles, with colored markers matching panels (l--n).
        The color scale applies to all oxygen maps, while the temperature scale applies only to panel (k).
    }
    \label{fig:sim_maps}
\end{figure*}

To connect the experimentally observed resistance changes and optical signatures to the underlying oxygen redistribution, we performed finite-element simulations using the framework described in Sec.~\ref{sec:methods_sim}. The simulated geometry was extracted from a SEM image of the device, thereby retaining fabrication-induced features such as slight terminal asymmetries and rounded corners in the central constriction. A close-up of the film geometry and mesh is shown in Fig.~\ref{fig:sim_maps}(a), while the full film-substrate simulation domain is shown in Fig.~\ref{fig:sim_maps}(f).

The simulation was initialized with a homogeneous oxygen content, $X_{\mathrm{O,init}} = 0.9$ (equivalent to 7-$\delta \approx 6.9$), and the same protocol as in the experiment was reproduced: At first, the left terminal was grounded, and a current $I_\mathrm{pulse}/2$ was injected through each of the two right-hand terminals, as sketched in Fig.~\ref{fig:sim_maps}(b). Upon increasing the pulse current, the simulated $R_\mathrm{min}(I_\mathrm{pulse})$ curve (Fig.~\ref{fig:sim_maps}(e)), measured between the left and top terminals, exhibits the same qualitative behavior as the experimental result presented in Fig.~\ref{fig:setup}(d): an initially weak variation at low pulse amplitudes, followed by a threshold-like increase of $R_\mathrm{min}$ with a rapidly increasing slope.

The corresponding oxygen maps in Fig.~\ref{fig:sim_maps}(b--d) reveal the microscopic origin of this resistance increase: Below the onset of EM, the oxygen landscape remains nearly unchanged (Fig.~\ref{fig:sim_maps}(b)). Near threshold, a depleted region forms on the left side of the narrowest part of the constriction, accompanied by an oxygen-rich region on the opposite side (Fig.~\ref{fig:sim_maps}(c)). At larger pulse amplitudes, the depleted region develops into a propagating front that travels into the targeted terminal, producing a substantial increase in $R_\mathrm{min}$ (Fig.~\ref{fig:sim_maps}(d)). This evolution is also visible in the line profiles extracted along the central axis of the left terminal, shown in the inset of Fig.~\ref{fig:sim_maps}(e): both the amplitude of the oxygen depletion and the distance of the front from the central node increase with continued EM. This behavior is consistent with Nanoprobe X-ray Diffraction experiments performed on these devices. A crystal lattice expansion is associated with oxygen depletion, based on the established correlation between oxygen stoichiometry and the YBCO lattice parameter reported in Ref.~\cite{jorgensen_structural_1990}. A similar approach, employing the same methodology to relate lattice expansion to oxygen content in comparable devices, has been previously reported in Ref.~\cite{QuaglioGomes2026}. Details on the results obtained for multiterminal devices are provided in Supplementary Material S3.

Starting from the final oxygen configuration of Fig.~\ref{fig:sim_maps}(d), a second current sweep was performed, this time targeting the top-right terminal. The resistance was monitored between the two right-hand terminals, as indicated in Fig.~\ref{fig:sim_maps}(g). Again, the simulated $R_\mathrm{min}(I_\mathrm{pulse})$ curve (Fig.~\ref{fig:sim_maps}(j)) displays a threshold-like increase. Interestingly, the previously formed oxygen-depleted front in the left terminal was modified during this second sweep not targeting this terminal. Even at the relatively low pulse amplitude of $2.5~\mathrm{mA}$, the sharp oxygen profile created during the first sequence is significantly smoothed, see Fig.~\ref{fig:sim_maps}(g). This behavior can be attributed to the large vacancy concentration gradients left behind by the first EM step, together with Joule-heating-enhanced diffusion during subsequent pulsing of the upmost right terminal. The line profiles in the inset of Fig.~\ref{fig:sim_maps}(j) confirm this progressive homogenization of the oxygen distribution. Further increasing $I_\mathrm{pulse}$ only weakly affects the already electromigrated left terminal, but instead drives the formation of a second oxygen-depleted front propagating into the newly targeted terminal Fig.~\ref{fig:sim_maps}(h, j).

The same behavior was reproduced when the third terminal, bottom right, was targeted. In this case, the resistance was tracked between the left terminal, which was targeted first, and the bottom-right terminal. At intermediate currents, $R_\mathrm{min}$ initially decreases (Fig.~\ref{fig:sim_maps}(o)), consistent with continued smoothing of the oxygen profile in the previously modified left terminal. This interpretation is supported by the line profiles shown in the inset of Fig.~\ref{fig:sim_maps}(o). A similar decrease in resistance has been reported in Refs.~\cite{Marinkovic2023, egyenes2026effect} and explained by oxygen disorder. At higher pulse amplitudes, a third oxygen-depleted front forms and progressively propagates outward along the newly targeted terminal (Fig.~\ref{fig:sim_maps}(l--n)). An animation with the full set of simulation frames is given in Supplementary Video 2. The temperature map in Fig.~\ref{fig:sim_maps}(k), evaluated at the end of the pulse leading to Fig.~\ref{fig:sim_maps}(n), shows that heating is strongly localized in the targeted terminal, whereas the two non-targeted branches, which each carry only half of the total current, remain at substantially lower temperatures. 

These numerical results provide useful insights on the respective roles of current flow and heating during the EM process. Local Joule heating is essential because it enhances oxygen mobility and promotes the relaxation of steep concentration gradients. However, the direction and location of the propagating oxygen-depleted fronts are dictated by the electrical configuration. The simulations therefore support an EM-dominated interpretation, with heating acting primarily as an enabling factor for oxygen diffusion rather than as the main directional driving force. The simulated oxygen maps reproduce several key features observed experimentally in the optical images of Fig.~\ref{fig:optical}. In both cases, the targeted terminal develops an outward-propagating oxygen-depleted front, while previously targeted terminals progressively lose contrast as their oxygen profiles relax. The simulations additionally predict an oxygen-rich region left behind the propagating depleted front. Interestingly, the local electric field extracted from the simulations reaches values on the order of $10^{6}\,\mathrm{V\,m^{-1}}$, slightly larger than those reported for $c$-axis carrier modulation induced by gate-controlled oxygen diffusion.\cite{Palau2018}

\begin{figure}[ht]
    \includegraphics[width=\linewidth]{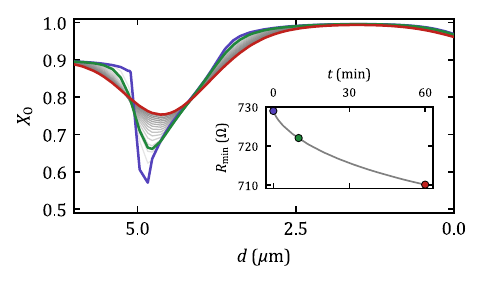}
    \caption{
        \textbf{Numerical relaxation starting from the oxygen configuration shown in Fig.~\ref{fig:sim_maps}(d): }
        The main panel shows the simulated oxygen content $X_\mathrm{O}$ along the indicated line profile at successive times during relaxation, plotted as a function of distance $d$ to the center of the device.
        The initially sharp oxygen-depleted region progressively broadens and partially refills, while the corresponding minimum resistance $R_\mathrm{min}$ decreases over the same relaxation window, as shown in the inset.
        Colored markers in the inset indicate the relaxation times of the highlighted profiles in the main panel.
    }
    \label{fig:sim_relax}
\end{figure}

To further isolate the relaxation mechanism, we performed a second type of simulation in which the oxygen configuration obtained after EM of the first terminal was allowed to relax for $1~\mathrm{h}$ without electrical stress. In this case, EM was disabled and the system evolved only through concentration-driven diffusion. The resulting evolution is shown in Fig.~\ref{fig:sim_relax}, where the main panel shows that the initially sharp oxygen-depleted region broadens and partially refills over time whereas the inset shows that the corresponding resistance decreases monotonically. The relaxation at 300~K is well described by a stretched exponential with $\tau = 81~\mathrm{min}$ and $\beta = 0.7$, in qualitative agreement with the experimentally observed relaxation behavior (Fig.~\ref{fig:EMs}(e)).

To summarize, the numerical implementation qualitatively captures the main trends observed in the different measurements: the threshold-like increase of $R_\mathrm{min}$ during EM, the selective formation of oxygen-depleted fronts in the targeted terminal, the smoothing of previously generated oxygen gradients during subsequent pulsing, and the slow relaxation of the resistance after electrical stress. In particular, the simulations show that the initial resistance decrease during subsequent terminal-targeting steps results from the partial reoxygenation of regions depleted during earlier EM runs. They also reveal that previously electromigrated regions are not static, but continue to evolve during later pulsing steps through diffusion driven by the concentration gradients left behind. The temperature maps further show why the process remains selective, since the largest heat-up is confined to the branch carrying the full pulse current. However, an exact quantitative agreement is not expected, given the strong sensitivity of YBCO transport and oxygen mobility to growth conditions, oxygen stoichiometry, local disorder, and possible deviations from the literature-reported transport parameters used in the simulation.

\section{Conclusion}
Using three-terminal YBCO devices, we demonstrated that electromigration can be applied to selectively control oxygen vacancy distribution, allowing modifications to be confined to a single terminal without significantly affecting other regions. This study provided clear evidence of directional effects of current-driven atomic migration, with current polarity playing a key role. We observed that a relaxation process driven by the vacancy concentration gradient occurs after EM even in the absence of electrical stress. The determined activation energy for relaxation $E_a = 0.52 \pm 0.01$~eV is consistent with established values for lattice oxygen diffusion in YBCO, supporting a diffusion-driven recovery mechanism. Oxygen redistribution was detected through changes in optical reflectivity, localized work-function variations measured by KPFM, and spatially resolved superconducting response obtained by SLM, which enabled quantitative mapping of oxygen content from approximately 6.50 in the targeted terminal to 6.85 in pristine regions. Finite-element simulations qualitatively reproduced the main experimental findings, providing deeper insight into the underlying dynamics of vacancy redistribution and relaxation.

These results establish controlled electromigration as a post-fabrication approach for branch-selective tuning of superconducting properties in multiterminal YBCO devices. This approach could further enable electrically defined oxygen-stoichiometry profiles for device trimming and control of current pathways in correlated-oxide systems without additional lithographic processing.

\section*{Acknowledgments}
The authors acknowledge financial support from the European Union’s Horizon 2020 research and innovation programme under Grant Agreement No. 101007417. This work benefited from access to the facilities of ICN2 – Fundació Institut Català de Nanociència i Nanotecnologia, Barcelona, Spain, through the NFFA-Europe Pilot Transnational Access Activity (Proposal ID 557).
This work was further supported by the F.R.S.-FNRS through the Excellence of Science (EOS) programme (Project No. O.0028.22) and by COST (European Cooperation in Science and Technology) through COST Action SUPERQUMAP (CA21144). D.S. and N.L. acknowledge support from the F.R.S.-FNRS through FRIA research fellowships.
C.C.Q.G. and M.M. acknowledge financial support from the São Paulo Research Foundation (FAPESP; Grants No. 2023/11915-1, 2025/03723-0, and 2022/03124-1) and the Brazilian National Council for Scientific and Technological Development (CNPq; Grant No. 310514/2025-8). C.C.Q.G., P.S., and M.M. also acknowledge support from the INCT Advanced Quantum Materials project, funded by CNPq (Processes No. 408766/2024-7 and 302786/2025-2), FAPESP (Process No. 2025/27091-3), and CAPES.
H.L., L.F., and A.P. acknowledge financial support from the Spanish Ministry of Science and Innovation through MCIN/AEI/10.13039/501100011033, under the “Severo Ochoa” Programme for Centres of Excellence (CEX2023-001263-S), the HTSUPERFUN project (PID2021-124680OB-I00), and the HTS-4ICT project (PID2024-156025OB-I00), co-funded by the European Regional Development Fund (ERDF), “A Way of Making Europe.” They also acknowledge support from the Spanish Nanolito networking project (RED2022-134096-T).
This research used facilities of the Brazilian Synchrotron Light Laboratory (LNLS), part of the Brazilian Center for Research in Energy and Materials (CNPEM), a private non-profit organization under the supervision of the Brazilian Ministry of Science, Technology and Innovation (MCTI). The authors acknowledge the CARNAÚBA beamline staff for their assistance during the experiments conducted under Proposal No. 20253173. 
The funding sources had no role in the design of the study; in the
collection, analysis, or interpretation of data; in the writing of the manuscript; or in the decision to submit the article for publication.
The authors gratefully acknowledge A. Baret and F. Balty for providing equipment and sharing their expertise in the automation of the experimental setup.

\section*{CRediT authorship contribution statement}
\textbf{Conceptualization:} A.V.S.
\textbf{Data curation:} D.S., C.C.Q.G.
\textbf{Formal analysis:} D.S., C.C.Q.G., A.V.S.
\textbf{Funding acquisition:} A.P., M.M., A.V.S.
\textbf{Investigation:} D.S., C.C.Q.G., N.L., E.F., P.S., H.L., L.F., A.P., M.M., A.V.S.
\textbf{Methodology:} D.S., C.C.Q.G., N.L., E.F., P.S.
\textbf{Project administration:} A.P., M.M., A.V.S.
\textbf{Software:} D.S., C.C.Q.G., N.L.
\textbf{Supervision:} M.M., A.V.S.
\textbf{Validation:} D.S., C.C.Q.G.
\textbf{Visualization:} C.C.Q.G., D.S.
\textbf{Writing -- original draft:} C.C.Q.G., D.S., A.V.S.
\textbf{Writing -- review \& editing:} D.S., C.C.Q.G., N.L., P.S., A.P., M.M., A.V.S.

\section*{Declaration of competing interest}
The authors declare that they have no known competing financial interests
or personal relationships that could have appeared to influence the work
reported in this paper.

\section*{Data availability} 
The main data supporting the findings of this study are available through the Universit\'e de Li\`ege Dataverse repository. Additional data are available from the corresponding authors upon reasonable request.

\section*{Declaration of generative AI and AI-assisted technologies in the manuscript preparation process}
During the preparation of this work, the authors used ChatGPT (OpenAI) to assist with language editing, improving clarity and structure, and refining the presentation and framing of selected sections of the manuscript. The authors critically reviewed, verified, and edited all AI-assisted content and take full responsibility for the content of the published article.

\section*{Supplementary material} 
Supplementary material to this article can be found online and includes additional results on the history dependence of the post-EM relaxation, experimental and numerical results for EM under positive current polarity, and nanoprobe X-ray diffraction measurements of the oxygen redistribution. Supplementary Video~1 shows the complete optical microscopy sequence during successive EM of the three terminals, while Supplementary Videos~2 and~3 show the simulated evolution of the oxygen distribution for negative and positive current polarities, respectively, including the subsequent relaxation in the positive-polarity case.

\bibliographystyle{elsarticle-num}
\bibliography{refs}

\end{document}


\begin{frontmatter}

\title{Electrical manipulation of oxygen stoichiometry in multiterminal
\texorpdfstring{YBa$_2$Cu$_3$O$_{7-\delta}$}{YBa2Cu3O7-delta} junctions}

\author[uliege]{Daniel Stoffels\corref{cor1}\fnref{equal}}
\ead{Daniel.Stoffels@uliege.be}

\author[uliege,ufscar]{Caio C. Quaglio-Gomes\corref{cor1}\fnref{equal}}
\ead{caioquaglio@estudante.ufscar.br}

\author[uliege]{Nicolas Lejeune}

\author[uliege]{Emile Fourneau}

\author[lnnano]{Pedro Schio}

\author[icmab]{Huidong Li}

\author[icmab]{Lourdes Fabrega}

\author[icmab]{Anna Palau}

\author[ufscar]{Maycon Motta}

\author[uliege]{Alejandro V. Silhanek\corref{cor1}}
\ead{asilhanek@uliege.be}

\cortext[cor1]{Corresponding authors.}

\fntext[equal]{These authors contributed equally to this work.}

\affiliation[uliege]{
    organization={Experimental Physics of Nanostructured Materials, Q-MAT,
    Department of Physics, Universit\'e de Li\`ege},
    city={Li\`ege},
    postcode={B-4000},
    country={Belgium}
}

\affiliation[ufscar]{
    organization={Departamento de F\'isica, Universidade Federal de S\~ao Carlos},
    city={S\~ao Carlos},
    postcode={13565-905},
    state={SP},
    country={Brazil}
}

\affiliation[lnnano]{
    organization={Brazilian Nanotechnology National Laboratory,
    Brazilian Center for Research in Energy and Materials},
    city={Campinas},
    postcode={13083-100},
    state={SP},
    country={Brazil}
}

\affiliation[icmab]{
    organization={Institut de Ci\`encia de Materials de Barcelona, ICMAB-CSIC},
    addressline={Campus UAB},
    city={Bellaterra},
    postcode={08193},
    country={Spain}
}

\end{frontmatter}

\renewcommand{\thesection}{S\arabic{section}}
\renewcommand{\thefigure}{S\arabic{figure}}
\renewcommand{\thetable}{S\arabic{table}}
\renewcommand{\theequation}{S\arabic{equation}}

\setcounter{section}{0}
\setcounter{figure}{0}
\setcounter{table}{0}
\setcounter{equation}{0}

\section{Invariance of relaxation dynamics to electromigration history}
\begin{figure*}[ht]
    \centering
    \includegraphics[width=0.95\linewidth]{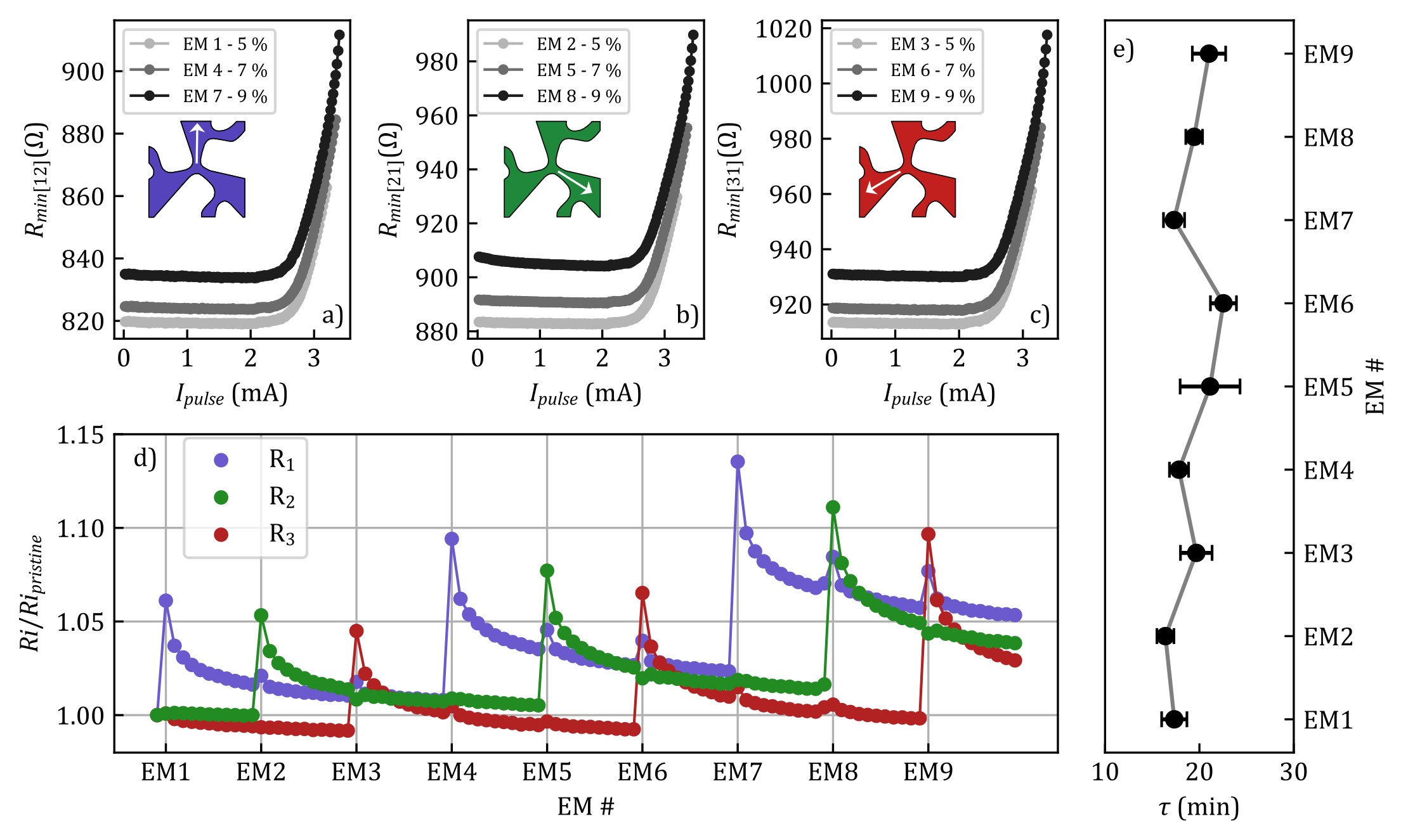}
    \caption{
        \textbf{Increasing cutoff criterion sweep:}
        $R_{\mathrm{min}}$ curves measured during a sequence of room temperature EM runs with increasing cutoff resistance on (a) R$_{1}$, (b) R$_{2}$, and (c) R$_{3}$. Panel (d) shows the evolution of individual resistances normalized by the pristine state resistances, with relaxation measured between the EM runs. Panel (e) summarizes the time constants $\tau$ obtained from exponential fittings of the resistance relaxation after each EM run. Notably, the results show good reproducibility despite differences in the EM cutoff and sample history, yielding a mean value of $\tau = 19.2 \pm 1.9$ min.
    }
    \label{fig:relax}
\end{figure*}
The dependence of the relaxation dynamics on the electromigration (EM) history was investigated using a protocol with progressively increasing cutoff resistance. One EM run was performed per terminal in the negative polarity, with a one-hour interval between consecutive runs. The cutoff resistance was initially set to 5\% above its initial value, and after all three terminals reached this threshold, it was increased in steps of 2\%. Resistance increases of up to 23\% (not shown) were achieved in a single EM run without device failure, demonstrating the large operational window of the devices. Figures~\ref{fig:relax}(a--c) show the resistance as a function of pulse current ($I_{\rm pulse}$) for the 5\% (EM1--EM3), 7\% (EM4--EM6), and 9\% (EM7--EM9) cutoff levels, respectively. Figure~\ref{fig:relax}(d) presents the normalized resistance measured immediately after each EM run and during the subsequent relaxation intervals. Stretched-exponential fits to the relaxation curves yielded the nine characteristic time constants shown in Fig.~\ref{fig:relax}(e), with an average value of $\tau = 19.2$~min and a standard deviation of 1.9~min. The absence of any systematic variation in $\tau$ with increasing cutoff resistance or successive EM runs indicates that, within the investigated range, the relaxation dynamics are independent of the EM history of the sample.

\section{Positive-polarity electromigration}
\begin{figure*}[ht]
    \centering
    \includegraphics[width=0.95\linewidth]{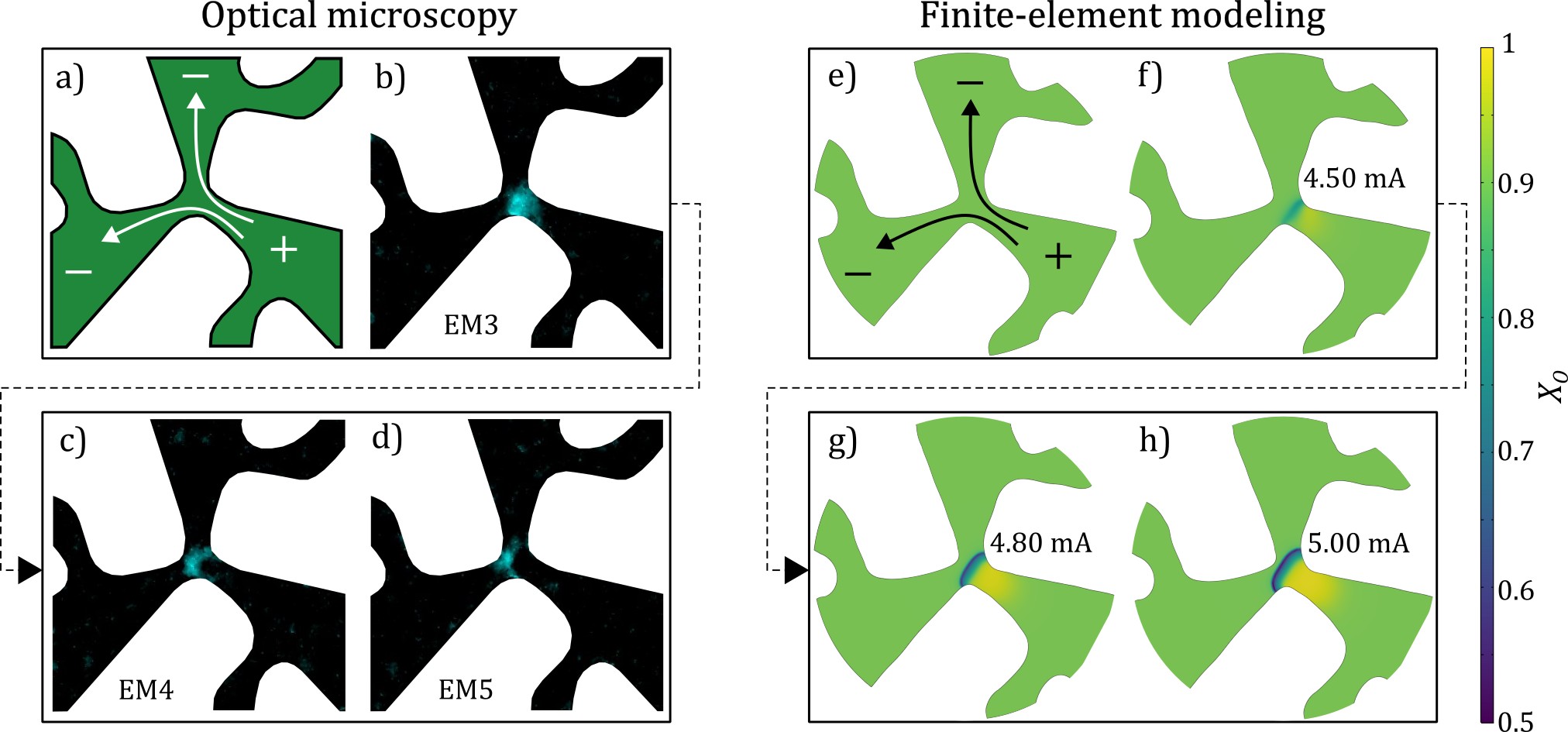}
    \caption{
        \textbf{Positive-polarity electromigration:} (a) Schematic of the electrical configuration and sequence of differential optical microscopy images acquired after (b) 3, (c) 4, and (d) 5 EM runs. (e) Electrical boundary conditions used in the finite-element model. Oxygen content ($X_O$) maps obtained under applied currents of (f) 4.50, (g) 4.80, and (h) 5.00 mA.
    }
    \label{fig:supfig2_positiveEM}
\end{figure*}

Figure~\ref{fig:supfig2_positiveEM} presents the evolution of the oxygen distribution during positive-polarity EM. The electrical configuration employed in these experiments is illustrated schematically in Fig.~\ref{fig:supfig2_positiveEM}(a), where the target terminal acts as the current source. Differential optical microscopy images acquired after 3, 4, and 5 electromigration runs are shown in Fig.~\ref{fig:supfig2_positiveEM}(b--d). The false-colored differential images reveal the progressive displacement of the oxygen-vacancy front toward the central node, highlighted by the blue contrast that expands with successive EM runs. This behavior provides direct experimental evidence that oxygen transport is governed by the direction of the applied current and that vacancy accumulation advances progressively into the central region of the device.

To further support this interpretation, Fig.~\ref{fig:supfig2_positiveEM}(e) defines the electrical boundary conditions employed in the finite-element simulations, while Fig.~\ref{fig:supfig2_positiveEM}(f--h) presents the corresponding oxygen-content ($X_O$) maps obtained under applied currents of 4.50, 4.80, and 5.00~mA, respectively. The simulations reproduce the main experimental features, showing the formation of an oxygen-depleted region on the inner side of the targeted junction, accompanied by oxygen accumulation on the opposite side. As the current amplitude increases, the depleted region evolves into a well-defined propagating front that advances toward the central node, in good qualitative agreement with the optical microscopy observations. A full animation of the positive polarity simulation profiles, consisting of an electromigration run followed by a relaxation for each terminal, can be found in Supplementary Video 3.

\section{Structural evolution profiling: X-ray diffraction}
Nanoprobe X-ray diffraction measurements at the CARNAÚBA beamline of the Sirius synchrotron facility (LNLS/CNPEM) were used to probe local structural changes induced by electromigration. Diffraction patterns were collected around the YBCO(005) reflection peak at multiple positions along the device, and analyzed by extracting the intensity-weighted center of diffraction peaks in detector space, followed by conversion to scattering vector. This procedure yields spatially resolved profiles of the out-of-plane lattice parameter ($c$-axis) along the direction from the device center toward the targeted terminal. A similar approach has been previously employed to extract spatially resolved lattice parameter variations induced by EM in related YBCO devices \cite{QuaglioGomes2026}.

\begin{figure*}[ht]
    \centering
    \includegraphics[width=0.95\linewidth]{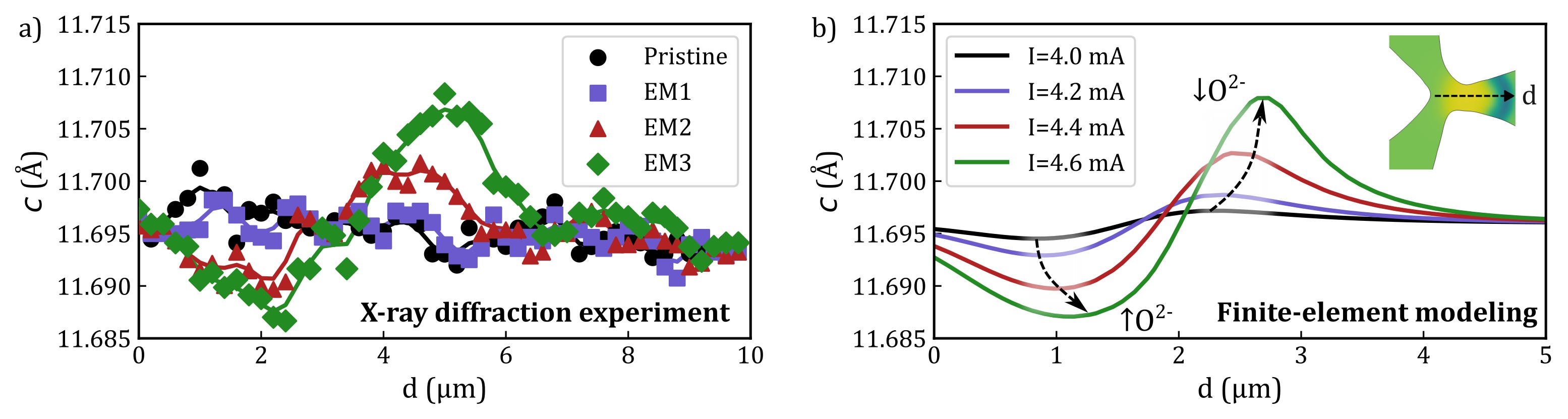}
    \caption{
        \textbf{Structural evolution profiling:} (a) X-ray diffraction line profiles along a terminal showing the measured $c$-lattice parameter in the pristine state (black circles) and after 1 (blue squares), 2 (red triangles), and 3 (green diamonds) EM runs. (b) Corresponding profiles extracted from oxygen-content maps obtained with the finite-element model under applied currents of 4.0 mA (black), 4.2 mA (blue), 4.4 mA (red), and 4.6 mA (green), and converted into $c$-lattice parameter profiles using a linear interpolation. The inset in (b) illustrates how the distance d is measured from the center of the device in both the experimental and numerical profiles.
    }
    \label{fig:supfig3_profiles}
\end{figure*}

Figure~\ref{fig:supfig3_profiles}(a) shows the experimentally measured $c$-axis line profiles in the pristine state and after 1--3 EM runs. Experimental symbols correspond to raw extracted values, whereas solid curves are used to highlight the underlying trends. A constant offset was subtracted between experimental scans to correct for alignment variations and instabilities in the beamline, enforcing consistency of the lattice parameter in the pristine region far from the device center. The uncertainty between scans was estimated as 0.0017~\AA~from the standard deviation of these offsets.

Because the $c$-axis lattice parameter is sensitive to oxygen content in the Cu--O chains of YBCO, these profiles directly reflect local oxygen redistribution. Along the device, a clear modulation develops with a peak in $c$ adjacent to a valley, consistent with oxygen depletion and accumulation regions, respectively. This feature grows with increasing number of EM runs, indicating progressive evolution of the underlying oxygen-vacancy distribution.

Finite-element simulations are presented in Fig.~\ref{fig:supfig3_profiles}(b) for selected current amplitudes of 4.0, 4.2, 4.4, and 4.6~mA. The inset illustrates the convention used to define the distance d from the device center. Simulated profiles are obtained by converting oxygen-content maps ($X_O$) into $c$-axis values using a linear interpolation between YBa$_2$Cu$_3$O$_6$ ($c = 11.84$~\AA) and YBa$_2$Cu$_3$O$_7$ ($c = 11.68$~\AA) \cite{jorgensen_structural_1990}. The simulations reproduce the main experimental features, including the characteristic peak--valley modulation associated with oxygen depletion and accumulation. While the modulation amplitudes are comparable between experiment and simulation, the spatial extent of the feature differs, consistent with the different effective electromigration conditions in each case.

\bibliographystyle{elsarticle-num}
\bibliography{refs}